\documentclass[sigconf]{acmart}
\AtBeginDocument{%
	\providecommand\BibTeX{{%
			\normalfont B\kern-0.5em{\scshape i\kern-0.25em b}\kern-0.8em\TeX}}}

\setcopyright{acmcopyright}
\copyrightyear{2022}
\acmYear{2022}
\acmDOI{10.1145/3544902.3546251}

\acmConference[ESEM '22]{ACM / IEEE International Symposium on Empirical Software Engineering and Measurement (ESEM)}{September 19--23, 2022}{Helsinki, Finland}
\acmBooktitle{ACM / IEEE International Symposium on Empirical Software Engineering and Measurement (ESEM) (ESEM '22), September 19--23, 2022, Helsinki, Finland}
\acmPrice{15.00}
\acmISBN{978-1-4503-9427-7/22/09}

\usepackage{multirow, tabularx}
\usepackage{subfigure}

\begin{document}
	
	\title{MMF3: Neural Code Summarization Based on Multi-Modal Fine-Grained Feature Fusion}
	
	
	\author{Zheng Ma}
	\affiliation{%
		\institution{Shandong Normal University}
		\city{Jinan}
		\country{China}}
	\email{mazhengwork@163.com}
	
	\author{Yuexiu Gao}
	\affiliation{%
		\institution{Shandong Normal University}
		\city{Jinan}
		\country{China}}
	\email{yuexiugao818@gmail.com}
	
	\author{Lei Lyu}
	\affiliation{%
		\institution{Shandong Normal University}
		\city{Jinan}
		\country{China}}
	\email{lvlei@sdnu.edu.cn}
	
	\author{Chen Lyu}
	\authornote{Corresponding author. Email: lvchen@sdnu.edu.cn}
	\affiliation{%
		\institution{Shandong Normal University}
		\city{Jinan}
		\country{China}}
	\email{lvchen@sdnu.edu.cn}
	
	\begin{abstract}
		\textbf{Background:} Code summarization automatically generates the corresponding natural language descriptions according to the input code to characterize the function implemented by source code. Comprehensiveness of code representation is critical to code summarization task. However, most existing approaches typically use coarse-grained fusion methods to integrate multi-modal features. They generally represent different modalities of a piece of code, such as an Abstract Syntax Tree (AST) and a token sequence, as two embeddings and then fuse the two ones at the AST/code levels. Such a coarse integration makes it difficult to learn the correlations between fine-grained code elements across modalities effectively. \textbf{Aims:} This study intends to improve the model's prediction performance for high-quality code summarization by accurately aligning and fully fusing semantic and syntactic structure information of source code at node/token levels. \textbf{Method:} This paper proposes a \textbf{M}ulti-\textbf{M}odal \textbf{F}ine-grained \textbf{F}eature \textbf{F}usion approach (MMF3) for neural code summarization. The method uses the Transformer architecture. In particular, we introduce a novel fine-grained fusion method, which allows fine-grained fusion of multiple code modalities at the token and node levels. Specifically, we use this method to fuse information from both token and AST modalities and apply the fused features to code summarization. \textbf{Results:} We conduct experiments on one Java and one Python datasets, and evaluate generated summaries using four metrics. The results show that: 1) the performance of our model outperforms the current state-of-the-art models, and 2) the ablation experiments show that our proposed fine-grained fusion method can effectively improve the accuracy of generated summaries. \textbf{Conclusion:} MMF3 can mine the relationships between cross-modal elements and perform accurate fine-grained element-level alignment fusion accordingly. As a result, more clues can be provided to improve the accuracy of the generated code summaries.
	\end{abstract}

	
	
	
	\maketitle
	
	\section{Introduction}
	Program comprehension is crucial for software development and maintenance, and source code equipped with a text summarization describing its functionality can significantly minimize developers’ cognitive efforts in program comprehension \cite{corbi1989program}. Therefore, high-quality summaries are essential to program comprehension and maintenance \cite{woodfield1981effect,tenny1988program,xia2017measuring}. Source code summarization technology can  generate brief natural language descriptions for source code \cite{latoza2006maintaining}. There is a proliferation of automated methods for generating code summaries, which are now essential for software development and maintenance. This approach helps developers break away from the tedious task of writing comments for code and significantly facilitates the understanding and maintenance of code for novice programmers.
	
	Recently, research on neural code summarization has blossomed, influenced by a wave of deep learning research \cite{hu2018summarizing,guo2020multi,wei2020retrieve,xie2021exploiting}. In particular, inspired by Neural Machine Translation (NMT) from natural language processing, many studies have used NMT-based frameworks to implement code summarization techniques, which treat code summarization process as a process of NMT. NMT is generally a neural architecture based on the encoder-decoder model for learning the mapping of two languages. Similarly, the model for code summarization uses this architecture to learn the mapping between code snippets and their natural language descriptions. For example, Iyer et al. \cite{iyer2016summarizing} used a sequence to sequence model with an attention mechanism to encode  code token sequences, which generates summaries of the source code. Afterward, some researchers (e.g., Hu et al. \cite{hu2018deep,hu2020deep}, LeClair et al. \cite{leclair2019neural}, and Alon et al. \cite{alon2018code2seq}) realized that in addition to the semantic information of token sequences, the structural information of ASTs \cite{wan2018improving} is also essential for the characterization of code snippets, and thus initiated an exploration of how to better mine the AST-embodied structural knowledge for code summarization. Since the above studies usually use Recurrent Neural Networks (RNNs) as the basic unit of the model, which are mainly divided into Long Short Term Memory (LSTM) \cite{hochreiter1997long} and Gate Recurrent Unit (GRU) \cite{cho2014learning}, it is challenging to capture long dependencies for lengthy code sequences \cite{bengio1994learning}. For this reason, Ahmad et al. \cite{ahmad2020transformer} proposed a code summarization model based on Transformer \cite{vaswani2017attention}. It uses token sequences of codes as model inputs and addresses the long dependency problem by replacing RNN with the self-attention mechanism, thus improving the accuracy of code summarization. The ensuing approaches \cite{zhang2020retrieval,yang2021multi,yang2021comformer,9796312} further incorporate structural information from ASTs into Transformer-based code summarization models.
	
	\textbf{Limitation:} 
	Existing NMT-based techniques have demonstrated the effectiveness of token sequences and ASTs. This suggests that multi-modal data participation has become a requirement for enriching program representations for code summarization. However, fusing two modalities with very different structures is still challenging. A more straightforward method is to represent the syntactic and semantic modalities of source code as two 1 × n dimensional embedding vectors, and then aggregate them into a single vector using some fusion method. Finally, the aggregated vector representation is used to guide models for summary prediction.
	
	The traditional fusion method is often implemented by vector stitching. For example, Yang et al. \cite{yang2021multi} proposed to learn source code representations from two heterogeneous modalities of AST, i.e., the structure-based traversal (SBT) sequences and graphs. Then they decoded the encoded two representations using a self-attentive mechanism, respectively, and finally spliced the decoded features to generate summaries. This stitching way is essentially a coarse-grained fusion method, which makes it difficult to explore the association between modalities and learn the different importance of the role played by each modality in the code summarization task. More recently, Gao et al. \cite{9796312} proposed to use the cross-modality feature fusion method to fuse vectors representing code tokens and AST modalities with weights based on attention scores for summary generation. Compared with the spliced fusion, this method can effectively discriminate the importance of each modality for summaries to be generated. However, since the fused objects are a pair of high-level features, the elemental knowledge (token semantics, node attributes) contained in the corresponding modalities and the detailed associations between these fine-grained elements cannot be shown in the fusion, which can easily damage the information related to summary generation and thus affect the quality of the results.
	
	In summary, coarse-grained fusion in code summarization loses more code detail information and adversely affects the generation ability of model performance. Therefore, how to pay more attention to the elements of each mode in the fusion and explore the fine-grained association and mapping between them is a challenge that needs to be solved.

	\textbf{Solution:} To address this challenge, we propose a multi-modal fine-grained feature fusion approach for neural code summarization. Our approach treats tokens as atomic elements that constitute the code and use them as granularity to discover correlations between different modalities. 
	%
	%
	
	\begin{figure}[h]
		\centering
		\includegraphics[width=1\linewidth]{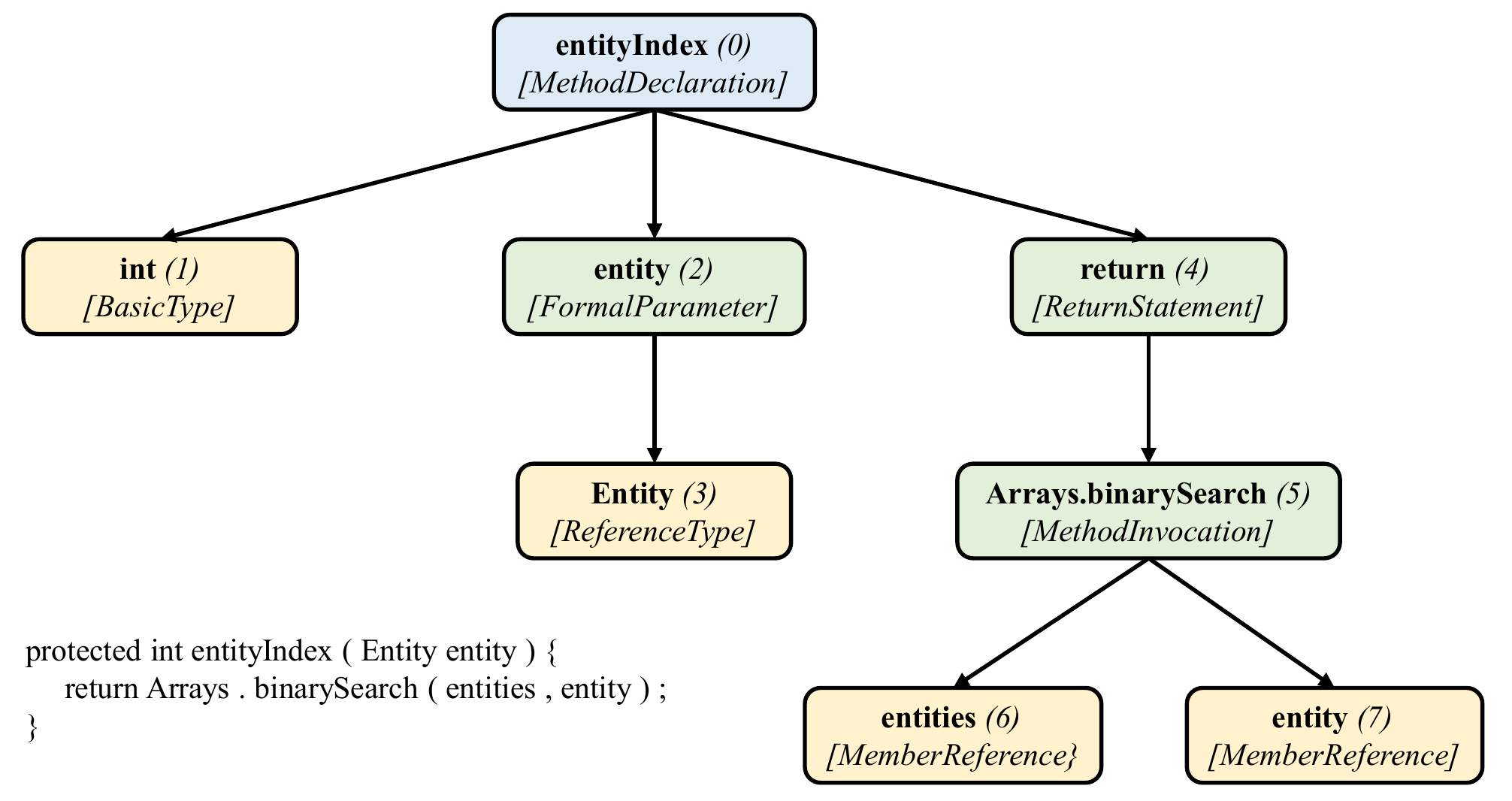}
		\caption{ A Java code and its AST}
		\Description{}
		\label{Fig:figure1}
	\end{figure}
	
	Specifically, we first characterize code snippets into two modalities, i.e., code token sequences and ASTs, and embed each element (token/node) of these two modalities using word embedding and Graph Convolutional Neural Network (GCN) to generate sequences embedding and ASTs embedding. Code comments are embedded as comments embedding by word embedding.

	Then, sequences embedding and ASTs embedding are encoded using the code encoder and AST encoder, respectively. Here the corresponding embedding vectors are generated and named as sequences feature and ASTs feature, which are combined together using a self-attention mechanism. After that, we present the fine-grained fusion method to fuse the there parts:
	combined feature, sequences embedding, and ASTs embedding. In the fusion, we determine the matching relationship between each token in token sequences and the value of each leaf node in ASTs by the strategy of comparing sameness in the order of position. Based on the obtained matching relationship, we then add the embedding vector of each leaf node of ASTs to the embedding vector of its matching token to achieve a fine-grained fusion of sequences embedding and ASTs embedding. Finally, sequences feature, fine-grained fused feature and comments embedding are fed into the decoder to train the entire model. Our approach allows fine-grained integration of token-level semantics of code snippets with node-level properties of ASTs. In this way, element-level tokens and nodes get adequate attention. Each modality's detailed semantics and structure are fully utilized by associating the corresponding elements to interact with each other in the fusion process to realize the complementary advantages of the modal features and thus can better serve the code summarization task.
	
	We conduct extensive experiments on two datasets, and experimental results demonstrate that MMF3 achieves better scores than current state-of-the-art methods on BLEU \cite{papineni2002bleu}, METEOR \cite{banerjee2005meteor}, ROUGE\_L \cite{lin2004rouge}, and CIDER \cite{vedantam2015cider} evaluation metrics. For example, on the Java dataset \cite{hu2018summarizing}, MMF3 scores 48.92\% and 59.94\% for BLEU-4 and ROUGE\_L, respectively, which are 2.05\% and 2.04\% higher than the current state-of-the-art method M2TS~\cite{9796312}, respectively.  MMF3 scores 35.92\% and 49.95\% for BLEU-4 and ROUGE\_L on the Python dataset \cite{barone2017parallel}, respectively, which are 2.04\% and 2.01\% higher than the current state-of-the-art method M2TS, respectively. In addition, we conduct extensive ablation experiments on each component of MMF3, and the results show that our proposed method benefits from each of the evaluated components.
	
	Our main contributions are summarized as follows:
	\begin{itemize}
		\item We propose a multi-modal fine-grained feature fusion method (MMF3) for neural code summarization, which can fuse multiple modal features of code snippets at a fine-grained level and use the fused information to improve the quality of code summarization.
		\item  We propose a multi-modal fine-grained feature matching method that can match tokens in a sequence of source code tokens with leaf nodes in an AST, enabling sophisticated correlation between multi-modal elements at the token and node granularity and facilitating a more comprehensive representation of the source code.
		\item We conduct a series of experiments to evaluate our approach on two real-word datasets, and the results confirm that MMF3 is effective and outperforms the state-of-the-art methods.
	\end{itemize}
	
	
	\section{BACKGROUND AND MOTIVATION}
	\subsection{Transformer-related Structures}
	The Transformer is a popular model in recent years, widely used in natural language processing and computer vision. It uses a complete Attention structure instead of LSTM and GRU, discarding the inherent pattern of the previous traditional encoder-decoder model that must be combined with Convolutional Neural Networks (CNNs) or RNNs. In addition, the Transformer can effectively reduce the computation and improve parallel efficiency. The following are details of several important structures used in the Transformer.
	
	\subsubsection{Positional Encoding:} Since the Transformer does not use the iterative operation of RNNs, we need to add the corresponding location information to each token input to the Transformer. For this purpose, we define a Positional Encoding concept. For example, the Transformer uses the traditional linear transformation of sin and cos functions to provide the model with location information.
	
	\begin{equation}
		PE(pos,2i) = \sin(pos/10000^{2i/d_{model}}),
	\end{equation}
	\begin{equation}
		PE(pos,2i+1) = \cos(pos/10000^{2i/d_{model}}),
	\end{equation}
	where pos refers to the position of a word in a sentence and takes values in the range [0,max\_sequence\_length) and $i$ refers to the dimensional ordinal number of the word vector and takes values in the range [0,embedding\_dimension/2), and d\_model in the formula refers to the value of embedding\_dimension.
	
	\subsubsection{Multi-head Attention:} This structure is an important part of the Transformer model. Self-attention defines three matrices $W^{Q}$, $W^{K}$, $W^{V}$, which are used to linearly transform all the token vectors input to the part to obtain the query matrix Q, the key matrix K, and the value matrix V, respectively. The difference between multi-head attention and self-attention is that the matrix of linear transformation is changed from one set $(W^{Q}, W^{K}, W^{V})$ to multiple sets $(W_j^{Q}, W_j^{K}, W_j^{V})$, and multiple sets $(Q_j,K_j,V_j)$ are obtained. Multi-head attention extracts different relationships between $(Q_j,K_j,V_j)$, and finally stitches the results together. The specific calculation formula is as follows:
	\begin{equation}
		Q_j = XW_j^{Q}, K_j=XW_j^{K}, V_j=XW_j^{V},
	\end{equation}
	\begin{equation}
		head_j = Attention(Q_j,K_j,V_j) = softmax(\frac{Q_jK_j^{T}}{\sqrt{d_K}})V_j,
	\end{equation}
	\begin{equation}
		MultiHead(Q,K,V) = Concat(head_1,\dots,head_n)W^{O},
	\end{equation}
	where $X$ is the input vector, $j$ denotes the $j\_th$ head, n denotes the total number of heads and $W^{O}$ denotes the linear transformation matrix.
	
	\subsubsection{Feed Forward Networks:} It is a two-layer neural network that first linearly transforms the input, then activates it with the activation function RELU, and finally linearly transforms it again. The specific calculation formula is as follows:
	\begin{equation}
		FFN(x) = max(0,xW_1 + b_1)W_2 + b_2,
	\end{equation}
	where $x$ is the input matrix, $W_1$ and $W_2$ are the weight matrices of each layer, $b_1$ and $b_2$ are their corresponding bias.
	
	\begin{figure*}[h]
		\centering
		\includegraphics[width=0.85\linewidth]{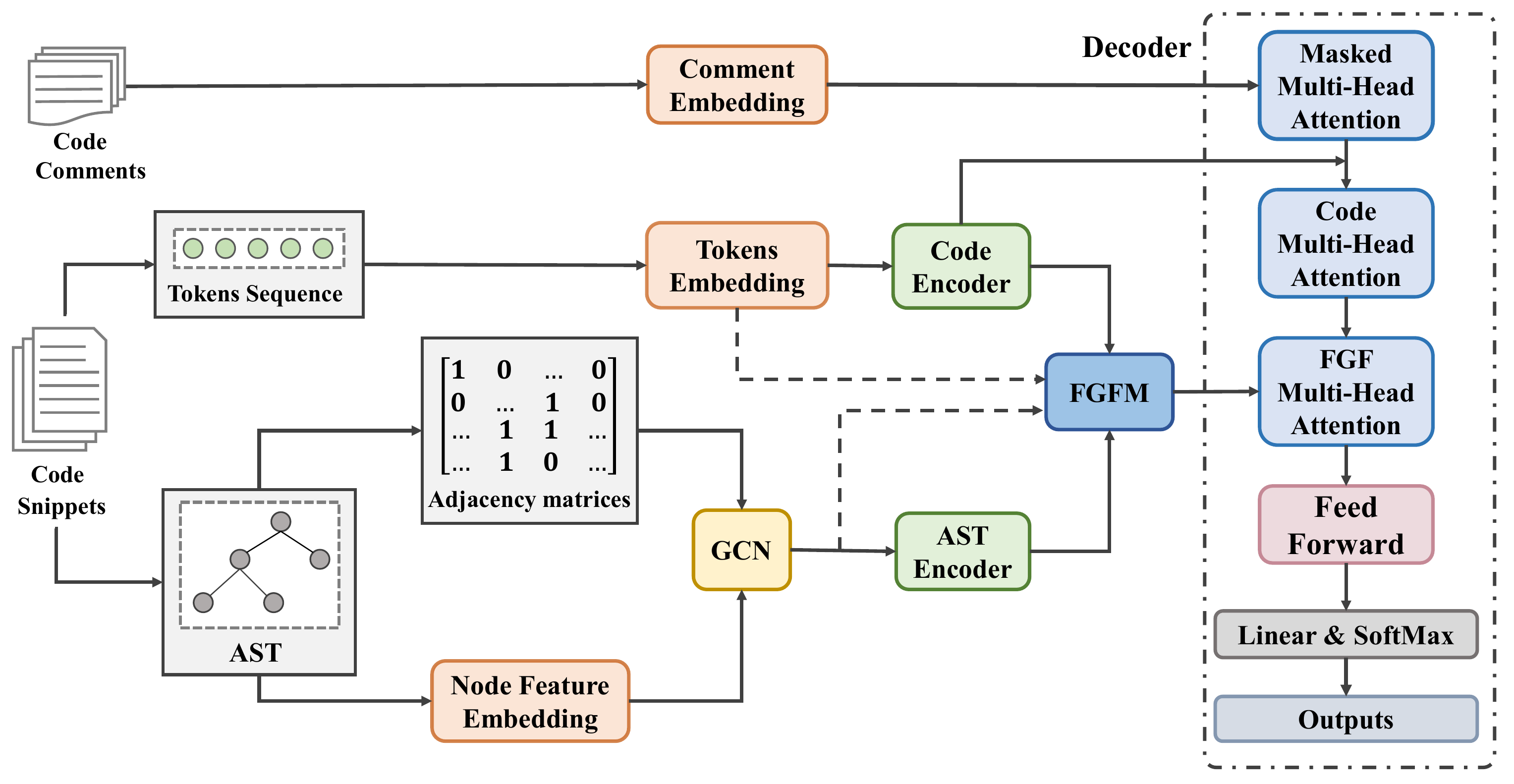}
		\caption{ The overall framework of our approach}
		\Description{}
		\label{Fig:figure2}
	\end{figure*}
	
	\subsection{Motivating Example}
	Most previous approaches use coarse-grained fusion methods to integrate information from multi-modal code snippets and perform AST-level or code-level vector fusion after representing ASTs or source code as vectors. In this way, the rich and precise semantic and syntactic information contained in fine-grained code elements (e.g., token and AST nodes) cannot be learned effectively. Figure \ref{Fig:figure1} shows a piece of Java method and the corresponding AST. Formally, the token modality and AST modality of this code fragments are represented as $X^{tok} = [x_1^{tok},x_2^{tok},x_3^{tok},\dots,x_m^{tok}]$,\break
	$X^{ast} =$ [$x_1^{ast},x_2^{ast},x_3^{ast},\dots,x_n^{ast}]$, the fusion of coarse-grained features is generally performed by 
	$X^{fus}{=}[x_1^{tok}{,}x_2^{tok},x_3^{tok},{\dots},x_m^{tok}$;
	$x_1^{ast},x_2^{ast},x_3^{ast},\dots,x_n^{ast}]$. Here $x_2^{tok}$ and $x_i^{ast}$ ($i$ is the position of the node in the AST corresponding to $x_2^{tok}$) are far apart and are still independent of each other. As a result, other modal-related information of the source code cannot be incorporated into the finer-grained token. To the best of our knowledge, the majority of  current code summarization models rely on encoder-decoder architectures, where models are trained to learn the mapping relationship between the source code and the summarization by inputting the source code and the corresponding summarization the encoder side and the decoder side, respectively \cite{vaswani2017attention,ahmad2020transformer}. Accordingly, the more precisely source code representation can be understood, the more accurately the model can learn the mapping. Therefore, a reasonable hypothesis is that if the model is trained by merging information about the AST modality of a code snippet into the embedding of the token associated with it, it can be made to understand the detailed connections across modalities more precisely, thus improving the model's understanding of multi-modal code representations. The challenge is to find the matching/fusion method that can integrate AST modalities into each token associated with them with precise alignment. To address this challenge, we conduct extensive research and analysis of the code as well as its corresponding AST, and observe that the values in the leaf nodes of the AST all have a token corresponding to them in the source code. Inspired by this phenomenon, MMF3 makes full use of this correspondence to explore the association between AST nodes and tokens for fine-grained multi-modal code representation fusion and code summarization. 
	
	\vspace{-0.4cm}
	\section{OUR APPROACH}
	Figure \ref{Fig:figure2} shows the overall framework of MMF3. Our proposed method consists of three main phases: data processing, MMF3 training, and MMF3 testing. We extract code snippets and corresponding natural language descriptions from both datasets. To comprehensively learn semantic and syntactic structural information of code snippets, we characterize them as token sequences and ASTs and use them as input to MMF3. MMF3 consists of two encoders (code encoder and AST encoder) and one decoder. Token sequences of the code snippets are input to the code encoder to learn semantic information of code snippets. ASTs are input to the AST encoder to learn syntactic structure information of code snippets. The outputs of the tokens embedding module and the GCN are sequences embedding and ASTs embedding, respectively. After that, in the training phase, sequences embedding, ASTs embedding, outputs of code encoder, and outputs of AST encoder are fed to the \textbf{F}ine-\textbf{G}rained \textbf{F}usion \textbf{M}odule (FGFM) to perform the fusion of token sequences and ASTs with fine-grained features. Then, code comments embedding, outputs of the code encoder, and outputs of FGFM are input to the decoder, where the pre-defined loss function are leveraged to optimize the whole network. In the testing phase, we input token sequences of code snippets and ASTs into our model, and the decoder generates the corresponding summaries based on the trained parameters. Note that, the architecture of the code encoder is the same as that of the Transformer. The critical parts of MMF3 model are mainly three modules: AST Encoder, FGFM, and Decoder, which are described below.
	
	\subsection{AST Encoder}
	In this paper, we use GCN to embed ASTs and empirically set the graph convolution layer in GCN to 2 \cite{leclair2020improved}. Initial feature embedding of the value of nodes of ASTs and adjacency matrix are input to the GCN layer, which outputs the embedding vector of ASTs. In particular, we use BERT pre-training method \cite{devlin2018bert} to  embed the node's value for each AST and do not involve it in the whole optimization process of the model. Finally, each node at the end of the GCN layer aggregates the adjacency information after the graph convolution. A multi-layer GCN with inter-layer propagation rules is formulated as follows:
	\begin{equation}
		H^{(l+1)} = \sigma(\hat{D}^{-\frac{1}{2}}\hat{A}\hat{D}^{-\frac{1}{2}}H^{(l)}W^{(l)}),
	\end{equation}
	where $\hat{A} = A + I_N$, $A$ denotes the adjacency matrix of the embedded graph,
	$I_N$ denotes the unit matrix, $\hat{D}_{ii} = \sum_{j}\hat{A}_{ij}$, and $H^{(l)}$ denotes the embedding vector of each node after $l$ GCN layers.
	
	According to the GCN propagation rule $f$, embedding vectors $X^{ast}$ of ASTs can be obtained from feature embedding $X^{node}$ of the value of nodes of ASTs and the adjacency matrix $Adj$.
	\begin{equation}
		X^{ast} = f(X^{node},Adj),
	\end{equation}
	
	\begin{figure}[h]
		\centering
		\includegraphics[width=\linewidth]{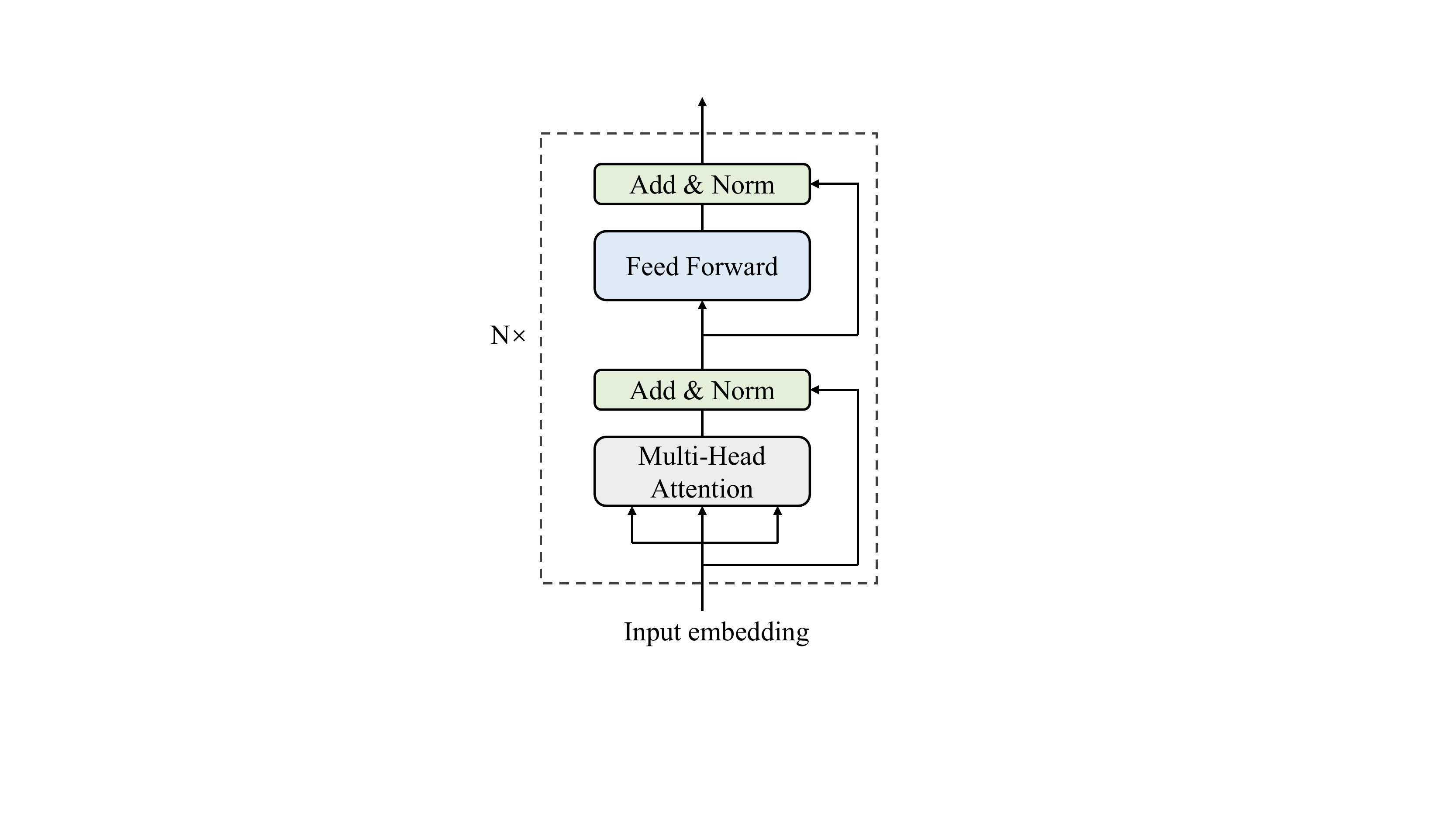}
		\caption{The structure of the AST encoder. $N$ indicates the number of layers}
		\Description{}
		\label{Fig:figure3}
	\end{figure}
	
	After that, the output $X^{ast}$ from the GCN module is fed to the AST encoder to further improve the ability to obtain grammatical information. Among them, the structure of the AST encoder is shown in Figure \ref{Fig:figure3}. It comes from a part of the Transformer architecture and consists of multi-head attention, layer normalization, and feed forward networks.

	\subsection{Fine-Grained Fusion Module}
	
\begin{figure}[h]
		\centering
		\includegraphics[width=\linewidth]{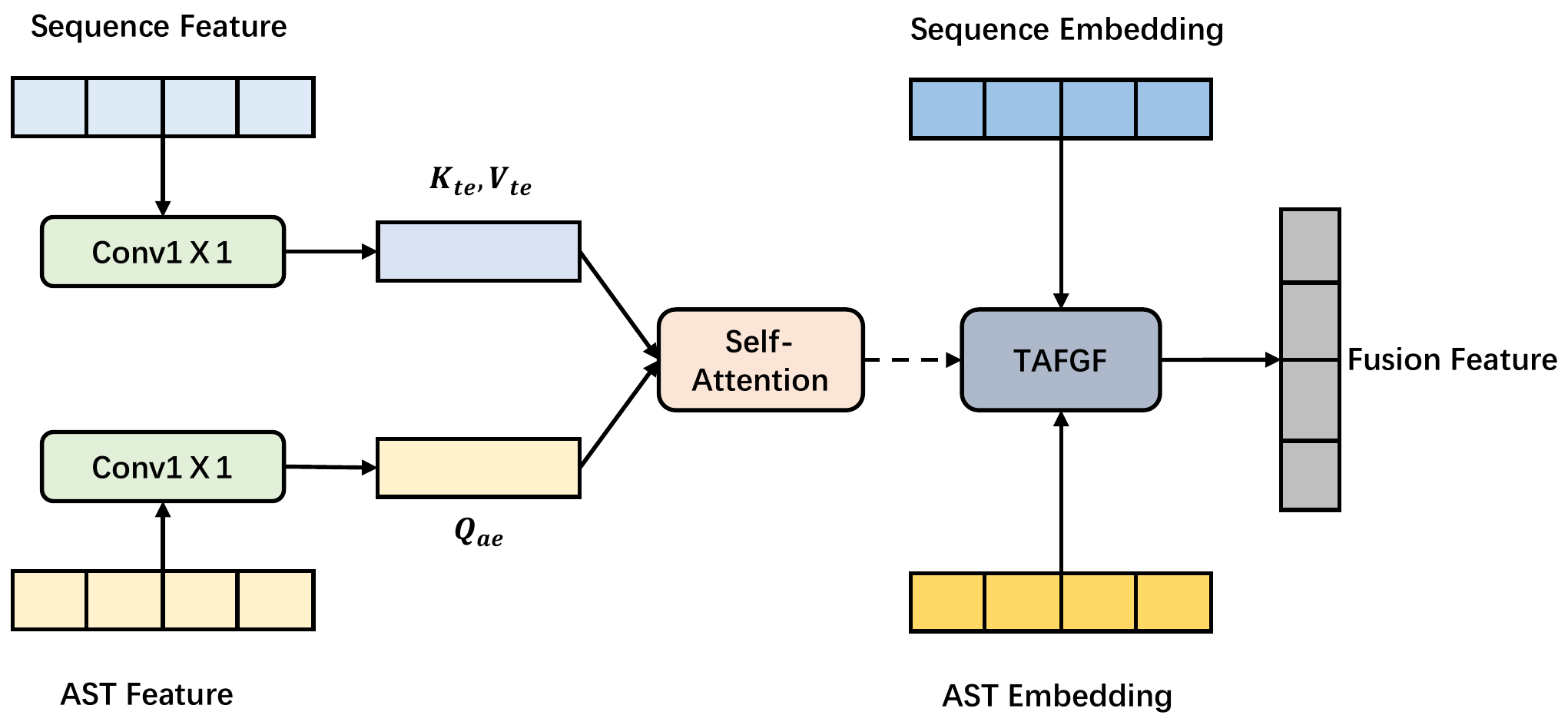}
		\caption{FGFM fusion process}
		\Description{}
		\label{Fig:figure4}
	\end{figure}

	This module is the core module of MMF3, and the overall process is shown in Figure \ref{Fig:figure4}. It uses a fine-grained fusion approach to fully fuse information from a token sequence and an AST of a code snippet, allowing the model to obtain a more comprehensive code feature. Specifically, the input of FGFM consists of four parts: sequence embedding, AST Embedding, the output of code encoder, and the output of AST encoder. The output of the code encoder is sequence feature (denoted as $X_e^{tok}$) and the output of the AST encoder is AST feature (denoted as $X_e^{ast}$). Among them, a combined code feature (denoted as $F1$) is obtained by combining the sequence feature and AST feature using the self-attention method, a fused code feature (denoted as $F2$) is obtained by fusing the sequence embedding and AST embedding using the fine-grained fusion method. Finally, $F$ is obtained by summing the two fused vectors, $F1$ and $F2$.
	
	Here we refer to the idea of the Transformer's self-attention mechanism to obtain $F1$, treating AST feature $X_e^{ast}$ as query and sequence feature $X_e^{tok}$ of the code as key and value. First, a 1 × 1 convolution of $X_e^{tok}$ and $X_e^{ast}$, with $X_e^{tok}$ convolved as $K_{te}$ and $V_{te}$ and $X_e^{ast}$ convolved as $Q_{ae}$, is calculated as follows, respectively.
	\begin{equation}
		Q_{ae} = Conv_{1\times1}(X_e^{ast}),
	\end{equation}
	\begin{equation}
		K_{te} = Conv_{1\times1}(X_e^{tok}),
		V_{te} = Conv_{1\times1}(X_e^{tok}),
	\end{equation}
	Then, the vector $F1$ after the fusion of sequence feature and AST feature is obtained after processing by the self-attention mechanism, and the specific formula is as follows:
	\begin{equation}
		F1 = softmax(\frac{Q_{ae}K_{te}^{T}}{\sqrt{d_k}})V_{te},
	\end{equation}
	where $d_k$ is set to 64.
	
	\begin{figure}[h]
		\centering
		\includegraphics[width=\linewidth]{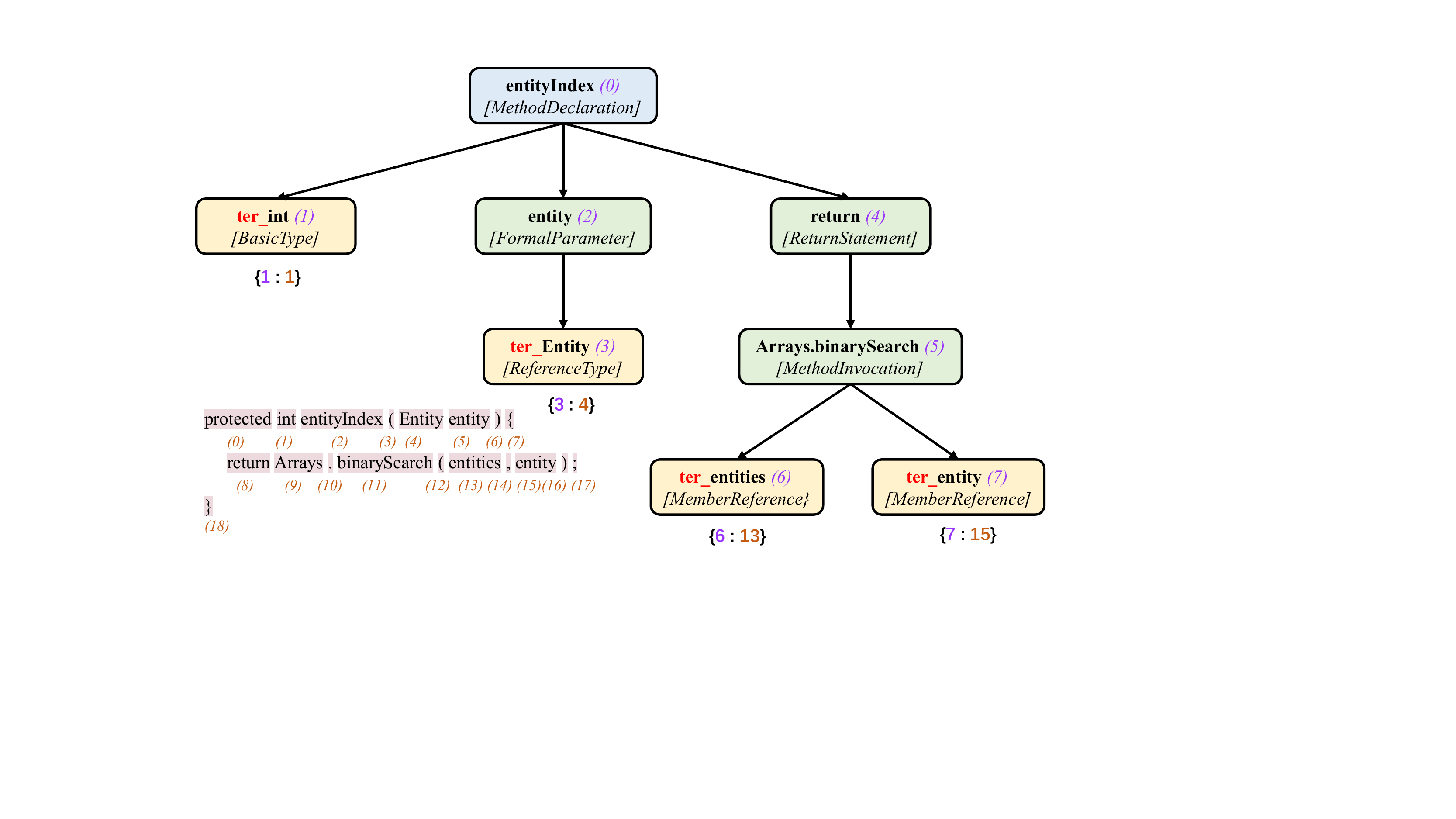}
		\caption{Fine-grained matching of Token modality and AST modality}
		\Description{}
		\label{Fig:figure5}
	\end{figure}
	
	To obtain $F2$, we design a sub-module in FGFM called SAFGF (standing for \textbf{F}ine-\textbf{G}rained \textbf{F}usion of the \textbf{S}equence embedding and the \textbf{A}ST embedding) to fine-grained fuse sequence embedding $X^{tok}$ with AST embedding $X^{ast}$ and add the embedding vector of leaf nodes of AST to the vector of matched tokens. This process is illustrated in Figure \ref{Fig:figure4}, where the purple numbers indicate the location number of the AST node and the brown numbers indicate the location number of the source code token. We identify the AST leaf node by adding the "ter\_" prefix to the value of the AST leaf node. Then, the value of the leaf node is compared with the token in the sequence of tokens in the order of position by string. If the value is the same as the token, then a mapping relationship is added between the leaf node's position and the token's position, which is stored in the form of a dictionary. Taking the code snippet and its AST in Figure \ref{Fig:figure5} as an example, "{3:4}" indicates that the value of the leaf node at AST position 3 is the same as the token at position 4 in the token sequence. At this point, a matching relationship is determined to exist between them. Thus, this form of dictionary building can achieve fine-grained matching between $X^{tok}$ and $X^{ast}$ of a code snippet. Finally, we add the embedding vector of the AST leaf node to its corresponding token vector, using $X^{tok}$ as the base. We adopt the newly obtained embedding vector of the sequence as the embedding vector $F2$ for fine-grained fusion, which is calculated as:
	\begin{equation}
		F2 = SUM(match(X^{tok},X^{ast})),
	\end{equation}
	
	
	\subsection{Decoder Module}
	The decoder module mainly converts the input token feature $X_e^{tok}$ and the FGFM output feature vector $F$ into the corresponding code summarization. During the training process, the module needs to input the corresponding comment embedding in addition to the above-mentioned $X_e^{tok}$ and $F$. The module consists of four main components: Masked Multi-Head Attention module, Code Multi-Head Attention module, FGF Multi-Head Attention module and Feed Forward module. As with the AST encoders used, a normalization is added after each block output in the decoder. 
	
	The input of the Code Multi-Head Attention module and the FGF Multi-Head Attention module is different with the Masked Multi-Head Attention module. The input Q, K, and V of the Masked Multi-Head Attention module are the embedding vector $Y_p$ after the sum of the comment embedding $Y$ and the corresponding positional embedding $P$. The output vector is the comment feature $Y_d$. The input Q, K, and V of the Code Multi-Head Attention module are $Y_d$, $X_d^{tok}$ and $X_d^{tok}$ respectively. The output is the vector $Y_d^{tok}$. The input Q, K, and V of the FGF Multi-Head Attention module are $Y_d^{tok}$, $F$ and $F$ respectively. The output is the vector $Y_d^{td}$. The Feed Forward module is the two layer fully connected , which input is $Y_d^{td}$ and output is the implicit vector $Y_{hidden}$. The cross-entropy loss function used in this paper is as:
	\begin{equation}
		Loss = -\frac{1}{n}\sum_{i=1}^{n}\sum_{j=1}^{m}logp(y_i^{(j)}),
	\end{equation}
	where $n$ denotes the total number of training sets, $m$ denotes the length of the generated target code summarization, $y_i^{(j)}$ denotes the j\_th word in the $i\_th$ sentence, and $p(y_i^{(j)})$ denotes the probability of generating the $j\_th$ word.
	
	In the training phase, the output vector of the code encoder and the fused vector of FGFM output need to be input into the Code Multi-Head Attention module and FGF Multi-Head Attention module, respectively, and the comment embedding needs to be input into Masked Multi-Head Attention module. In the testing phase, only the output vector of the code encoder and the fusion vector of FGFM output are required to be input into the Code Multi-Head Attention module and FGF Multi-Head Attention module, respectively.
	
	\section{EXPERIMENTAL SETUP}
	\subsection{Datasets and Preprocessing}
	We identify two commonly used datasets from \cite{9796312}, the Java dataset used by Hu et al \cite{hu2018summarizing}. and the Python dataset used by Barone et al \cite{barone2017parallel}. In this paper, we call these two datasets as JHD $\footnote{\url{https://github.com/xing-hu/TL-CodeSum}}$ and PBD $\footnote{\url{https://github.com/EdinburghNLP}}$. Specifically, we cite the 78422 pairs as the JHD dataset collected in the repositories created on GitHub. The PBD dataset we used contains 108726 pairs of python functions and their comments from the GitHub open source repositories. Since our approach requires AST modalities for code snippets, we use two tools for parsing code to ASTs: JDK compiler and Treelib toolkit, and parsing JHD and PBD, respectively. The statistics of these two datasets are described in Table \ref{Tab: table1}, where the AvgL and UnitL are the average length and the total number of unique tokens, respectively. The MaxN and AvgN denote the maximum and average number of nodes. Consistent with the baselines for comparison, we split PBD into the training set, validation set, and testing set with fractions of 60\%, 20\%, and 20\%, respectively, and split JHD in a proportion of 8: 1: 1. To make the training and testing sets disjoint, we remove the duplicated samples from the testing set \cite{zhang2020retrieval}. Since there are differences between the code snippets in these two datasets, we use the word segmentation approach of CamelCase and snake case for both datasets and built the respective data dictionaries (vocabularies). In addition, we use <PAD> to fill the data in the dataset to facilitate batch processing of the dataset so that the length of each piece of data is consistent. We also add <SOS> and <EOS> to target sequences as the start and end markers, respectively. Finally, we streamline the dataset by removing code snippets with empty summaries, as such data is not meaningful for the target task.
	\begin{table}
		\caption{Statistical analysis of the two datasets}
		\label{Tab: table1}
		\begin{tabular}{c c c c c c c}
			\toprule
			\multirow{2}{*}{Dataset} & \multicolumn{2}{c}{Code Length} & \multicolumn{2}{c}{NL Length} & \multicolumn{2}{c}{AST(Node)}\\
			\cmidrule(r){2-3} \cmidrule(r){4-5} \cmidrule(r){6-7}
			&AvgL  & UnitL &AvgL  & UnitL &MaxN  & AvgN\\
			\midrule
			JHD & 120.99 & 23501 & 17.71 & 29645 & 2126 & 50\\
			PBD & 57.12 & 125584 & 9.82 & 31116 & 1522 & 87  \\
			\bottomrule
		\end{tabular}
	\end{table}
	
	\subsection{Experimental Settings}
	The embedding dimension is 512, and the feedforward dimension is set to 2048 in our MMF3 model. The dimensions of Q, K, and V of the input multi-head attention are set to 64, the number of code encoder layers, AST encoder layers, and code decoder layers are all set to 6, and the number of heads in the multi-head attention is set to 8. We used an SGD optimizer to train MMF3 and set the learning rate to 0.001. We set the minimum batch size to 32, the dropout rate to 0.2, and used a beam search of size 4. We train M2TS for at most 200 epochs and adopt an early stop if the validation performance does not improve after 20 epochs. The experiments in this paper are all conducted on the Intel(R) Xeon(R) Gold 5218 CPU @ 2.30GHz, 128GB RAM, and TITAN RTX 24G GPU platform.
	
	\subsection{Evaluation Metrics}
	Similar to existing work \cite{wan2018improving,hu2018deep,hu2020deep,zhang2020retrieval,yang2021multi,9796312}, we use four evaluation metrics commonly used in machine translation and text summarization to evaluate the quality of the generated summaries: BLEU \cite{papineni2002bleu}, METEOR \cite{banerjee2005meteor}, ROUGE\_L \cite{lin2004rouge} and CIDER \cite{vedantam2015cider}.
	
	\textbf{BLEU} measures the n-gram precision between the generated summaries and the ground truth by computing the overlap ratios of n-grams and applying the brevity penalty on short translation hypotheses. The specific formula for BLEU-N is as follows:
	\begin{equation}
		BLEU-N = BP \cdot exp\sum\limits_{n=1}^N \omega_n\log p_n , 
	\end{equation}
	where $p_n$ is the precision score of the n-gram matches between candidate and reference sentences. BP is the brevity penalty and $\omega_n$ is the uniform weight $1/N$. This paper uses the BLEU-4 scores to evaluate the metric for generating code summaries.
	
	\textbf{METEOR} evaluates how well the generated comments capture content from the references via recall, computed by stemming and synonymy matching. It is calculated as:
	\begin{equation}
		METEOR = (1- \gamma\cdot frag^\beta) \cdot \frac{P \cdot R}{\alpha \cdot P + (1-\alpha) \cdot R},
	\end{equation}
	where the $P$ and $R$ are the precision and recall, the $frag$ is a fragmentation fraction. The default values of the parameters, $\gamma$, $\beta$, $\alpha$ are 0.5, 3.0 and 0.9, respectively.
	
	\textbf{ROUGE\_L} evaluates how much reference text appears in the generated text. Based on the longest common subsequence (LCS), it uses the F-score, which is the harmonic mean of precision and recall values. For example, suppose the lengths of $X$ and $Y$ are $m$ and $n$, then:
	\begin{equation}
		P_{lcs} = \frac{LCS(X, Y)}{m}, R_{lcs} = \frac{LCS(X, Y)}{n}, 
	\end{equation}
	\begin{equation}
		F_{lcs} = \frac{(1+\beta^2)P_L \cdot R_{lcs}}{R_{lcs} + \beta^2P_{lcs}},
	\end{equation}
	where the $\beta$ is set to 1.2 as in \cite{wan2018improving}, and the $F_{lcs}$ is the value of ROUGE\_L.
	
	\textbf{CIDER} is a consensus-based evaluation metric for measuring image captioning quality. $CIDER_n$ score for n-gram is computed using the average cosine similarity between the candidate and reference sentences.
	
	Note here that BLEU, METEOR, and ROUGE-L scores are in the $[0,1]$ range and are usually displayed as percentages. However, CIDER is not between 0 and 1, so it is displayed as an actual value.
	
	\subsection{Baselines}
	We compare MMF3 with existing approaches of source code summarization. A detailed description of the currently available methods for source code summarization is given below.
	
	\textbf{CODE-NN} is proposed by Iyer et al. \cite{iyer2016summarizing} and uses the LSTM to compose an end-to-end generative system. 
	CODE-NN initializes all model parameters uniformly between $-0.35$ and $0.35$ and starts with a learning rate of 0.5. 
	
	\textbf{Hybrid-DeepCom} uses the encoder-decoder framework to input the token sequence of code snippets and the AST sequence processed by the SBT method to the code encoder and AST encoder of the model, respectively \cite{hu2020deep}. 
	
	\textbf{Code2Seq} is a standard encoder-decoder framework proposed by Alon et al. \cite{alon2018code2seq}. The encoder side inputs a joint representation of the vector of each path of the code fragment AST with the token vector, and the decoder side outputs the corresponding summaries. 
	The values of all parameters of Code2Seq are initialized using the initialization heuristic of Glorot and Bengio \cite{glorot2010understanding}. 
	
	\textbf{Code+GNN+GRU} is proposed by LeClair et al. \cite{leclair2020improved}, which uses GNN to embed ASTs. They input embedded AST vectors and code token sequences embedding into the GRU, respectively. Then, code comments are output by the decoder with attention mechanism.
	
	\textbf{V-Transformer} replaces all RNNs and CNNs previously used in the encoder-decoder framework with the self-attention mechanism and effectively solves the long-term dependency problem \cite{vaswani2017attention}.
	
	\textbf{NeuralCodeSum} is proposed by Ahmad et al. \cite{ahmad2020transformer}, which adds copy attention and relative position encoding to the overall framework of the traditional Transformer. 
	
	\textbf{MMTrans} represents code snippets as ASTs and corresponding SBT sequences, and then input them to the graph encoder and SBT encoder, respectively \cite{yang2021multi}. A joint decoder generates the corresponding code summaries. 
	
	\textbf{SG-Trans} is proposed by Gao et al. \cite{gao2021code} to inject local semantic information and global syntactic structure into the self-attention module in the Transformer as an inductive bias better to capture the hierarchical features of the source code. SG-Trans uses the Adam optimizer and sets the initial learning rate too. 
	
	\textbf{ComFormer} uses the overall architecture of the Transformer model proposed by Yang et al. \cite{yang2021comformer}, where the sequence of tokens of the source code and the sequence of AST traversal by the Sim\_SBT method are input at the encoder side. 
	
	\textbf{SeCNN} is proposed by Li et al. \cite{li2021secnn}, and uses two layer CNN to encode semantic information of the source code. One CNN is used to extract lexical information from the code tokens, and another CNN is used to extract syntactic information from the AST. 
	
	\textbf{AST-Trans} aims to convert ASTs of code snippets into sequences by linearization method, then encode the ancestor-descendant and sibling relationships in the ASTs into two matrices, and finally, generate code summaries by the relationship matrix and self-attention mechanism \cite{tang2021ast}.
	
	\textbf{EDITSUM} is a new retrieval and editing method for code summaries, consisting of a retrieval module and an editing module \cite{li2021editsum}. 
	The model uses Adam as the optimizer, setting the learning rate to 0.001. 
	
	\textbf{CAST} can split and reconstruct the AST hierarchically \cite{shi2021cast}. Its main idea is to split a large AST hierarchically into a set of subtrees, and then aggregate the representations of the subtrees to reconstruct the split AST.
	
	\textbf{M2TS} uses a multi-scale approach to extract the structural features of ASTs and uses a cross-modal fusion method to fuse them with the token sequence information of source code \cite{9796312}. 
	M2TS uses the SGD optimizer and sets the initial learning rate to 0.0001. 

	\begin{table*}
		\caption{Comparison of our proposed approach with the baseline approaches}
		\label{Tab: table2}
		\begin{tabular}{lcccccccc}
			\toprule
			\multirow{2}{*}{\textbf{Approach}} & \multicolumn{4}{c}{JHD} & \multicolumn{4}{c}{PBD}\\
			\cmidrule(r){2-5} \cmidrule(r){6-9}
			& BLEU-4 & METEOR & ROUGE\_L & CIDER & BLEU-4 & METEOR & ROUGE\_L & CIDER\\
			\midrule
			CODE-NN & 26.58\% & 13.43\% & 40.60\% & 0.864 & 16.78\% & 9.27\% & 37.17\% & 1.008\\
			Hybrid-DeepCom & 38.64\% & 23.41\% & 51.47\% & 1.263 & 21.24\% & 9.83\% & 38.01\% & 1.303\\
			Code2Seq & 38.97\% & 23.85\% & 51.92\% & 1.372 & 22.05\% & 10.53\% & 38.32\% & 1.378\\
			Code+GNN+GRU & 40.36\% & 24.38\% & 52.90\% & 1.868 & 23.67\% & 12.50\% & 40.03\% & 1.786\\
			V-Transformer & 42.66\% & 25.43\% & 53.27\% & 2.053 & 30.78\% & 17.87\% & 43.25\% & 2.023\\
			NeuralCodeSum & 43.51\% & 26.13\% & 54.12\% & 2.044 & 31.26\% & 19.09\% & 45.69\% & 2.257\\
			MMTrans & 45.21\% & 27.01\% & 55.32\% & 2.357 & 32.56\% & 18.45\% & 46.46\% & 2.337\\
			SG-Trans & 45.33\% & 27.23\% & 55.87\% & 2.313 & 32.50\% & 19.77\% & 46.54\% & 2.252\\
			ComFormer & 46.05\% & 27.54\% & 55.84\% & 2.374 & 33.15\% & 21.14\% & 46.95\% & 2.323\\
			SeCNN & 45.23\% & 26.98\% & 55.24\% & 2.342 & 32.68\% & 19.25\% & 46.53\% & 2.326\\
			AST-Trans & 46.14\% & 27.93\% & 55.96\% & 2.372 & 33.32\% & 21.45\% & 47.26\% & 2.356\\
			EDITSUM & 46.32\% & 28.06\% & 56.13\% & 2.413 & 33.54\% & 21.62\% & 47.53\% & 2.385\\
			CAST & 46.57\% & 28.53\% & 56.62\% & 2.492 & 33.64\% & 21.68\% & 47.74\% & 2.396\\
			M2TS & 46.87\% & 28.98\% & 57.93\% & 2.578 & 33.88\% & 21.87\% & 47.97\% & 2.415\\
			MMF3 & \textbf{48.92\%} & \textbf{29.99\%} & \textbf{59.97\%} & \textbf{2.683} & \textbf{35.92\%} & \textbf{22.91\%} & \textbf{49.98\%} & \textbf{2.523}\\
			\bottomrule
		\end{tabular}
	\end{table*}
	
	\section{RESULTS AND ANALYSIS}
	Our study aims to assess whether MMF3 outperforms the latest baseline and why the multi-modal fine-grained feature fusion method can improve the quality of the generated summaries. In this section, we give the experimental results and analysis the proposed research problems.
	
	\subsection{RQ1: How does our approach perform compared to the baselines?}
	We conduct experiments on one Java and one Python datasets. %
	%
	In reproducing the baselines, we try to prepare the input following the data processing steps of the baselines and set the hyperparameters to be as consistent as possible with this paper, thus ensuring the fairness of experiments.
	The specific experimental results are shown in Table \ref{Tab: table2}. We can see that MMF3 obtains the best results, better than the current latest competitors. Overall, the reason for this result is that we fuse the fine-grained features of the embedding vectors of the code token sequences and ASTs. In this way, both the tokens in token sequences and the nodes in ASTs are given sufficient attention to achieve the complementary advantages between the modal features, thus characterizing the code information more comprehensively and serving summaries better. 
	%
	
	Specifically, MMF3 scores 22.34\%, 6.26\%, and 5.41\% higher than CODE-NN, V-Transformer, and NeuralCodeSum on the JHD dataset in terms of BLEU-4. The reason is that all these three models, CODE-NN, V-Transformer, and NeuralCodeSum, only use token sequences as input, ignoring the syntactic structure information of code snippets. V-Transformer and NeuralCodeSum are superior to CODE-NN because CODE-NN does not use the structure of the Transformer but uses an encoder-decoder framework composed of LSTM units, which is struggle to solve the long dependency problem. On the PBD dataset, MMF3 is 13.08\% and 1.77\% higher in METEOR than Hybrid-DeepCom and ComFormer, respectively. The main reason is that both Hybrid-DeepCom and ComFormer use token sequences of source code and the traversal sequences of ASTs to input the model to predict summaries, ignoring the fact that the traversal sequence of the AST is not able to express the structural information of the AST effectively. In addition, Hybrid-DeepCom and ComFormer use stitching for a fusion of embedding vectors, which lacks a fine-grained synthesis of the semantics of elements and their associations across modalities. ComFormer outperforms Hybrid-DeepCom since it uses a Transformer-based fusion method that better balances the weight of token and AST modalities in the fusion vector.
	
	On the JHD dataset, MMF3 is 2.04\% and 3.35\% higher in CIDER scores than M2TS and CAST, respectively. The reason is that MMF3 uses GCN to extract the AST information from the source code fully and to perform fine-grained fusion (between token-level and node-level fusion) of the Token modality of the source code with the AST modality. In addition, it is observed that M2TS significantly outperforms Code2Seq, Code+GNN+ GRU, and SeCNN. We analyze the reasons behind this result. M2TS uses a new cross-modality feature fusion method to a weighted fusion of vectors corresponding to token sequence modality of code snippets with vectors corresponding to AST modality based on attention scores, which is then used for summary generation. In contrast to stitching fusion, this approach allows the importance of each modality for the summarization to be generated to be effectively distinguished. SeCNN outperforms Code2Seq and Code+GNN+GRU because SeCNN convolves the traversal sequence of AST to obtain more comprehensive information about AST structure. On the JHD dataset, MMF3 is 3.84\% higher than EDITSUM on ROUGE\_L. This reason is mainly that EDITSUM only considers the semantic information of the source code and does not consider the structural information of the AST. On the PBD dataset, MMF3 is 0.186 and 0.167 higher in CIDER scores than MMTrans and AST-Trans, respectively. The reason is that both MMTrans and AST-Trans only explore the structural information of the AST modality of the source code and do not consider the information of the Token modality of the source code, the lack of which would compromise the integrity of the code representation. On the JHD dataset, MMF3 has a 2.76\% higher METEOR score than SG-Trans, which suggests that MMF3 can express the semantic and syntactic information of code snippets more completely than SG-Trans. Table \ref{Tab: table2} shows that the scores of the evaluation metrics for the various methods are higher on JHD than on PBD, which may be due to the different programming languages used in the datasets.
	
	\begin{table}
		\caption{Comparison of setting different heads on the JAH dataset}
		\label{Tab: table3}
		\resizebox{1\columnwidth}{!}{
			\begin{tabular}{lccccc}
				\toprule
				\textbf{Approach} &Head-size & BLEU-4 & METEOR & ROUGE\_L & CIDER \\
				\midrule
				\multirow{6}{*}{MMF3}
				& 2& 45.78\% & 28.80\% & 58.93\% & 2.567 \\
				& 4& 47.82\% & 29.03\% & 58.86\% & 2.623 \\
				& 6& 48.11\% & 29.30\% & 59.23\% & 2.615 \\
				& 8& \textbf{48.92\%} & \textbf{29.99\%} & \textbf{59.97\%} & \textbf{2.683} \\
				& 10& 47.95\% & 28.87\% & 58.90\% & 2.574 \\
				& 12& 46.89\% & 28.54\% & 58.76\% & 2.563 \\
				\bottomrule
			\end{tabular}
		}
	\end{table}

	\subsection{RQ2: How the multi-head attention mechanism heads affect the performance of MMF3?}
	MMF3 uses components from the Transformer architecture, the most important of which is the multi-head attention mechanism. Multi-head attention functions to assign corresponding weights to different input information and perform weighting calculations. Among them, the size of the head number setting will impact the experimental results. To this end, we specially design experiments to explore how many heads are set to make the model achieve the best experimental results. We set the number of heads as 2, 4, 6, 8, 10, and 12 and conduct experiments on the JHD dataset, respectively, and the experimental results are shown in Table \ref{Tab: table3}. It can be observed that the best experimental results were obtained when the number of heads is 8.
	
	\begin{table*}
		\caption{Ablation experiments on two datasets}
		\label{Tab: table4}
		\begin{tabular}{lcccccccc}
			\toprule
			\multirow{2}{*}{\textbf{Approach}} & \multicolumn{4}{c}{JHD} & \multicolumn{4}{c}{PBD}\\
			\cmidrule(r){2-5} \cmidrule(r){6-9}
			& BLEU-4 & METEOR & ROUGE\_L & CIDER & BLEU-4 & METEOR & ROUGE\_L & CIDER\\
			\midrule
			MMF3(AST) & 46.48\% & 27.92\% & 57.59\% & 2.562 & 33.37\% & 20.34\% & 47.43\% & 2.312\\
			MMF3(SAN) & 48.03\% & 29.32\% & 59.04\% & 2.624 & 34.97\% & 21.95\% & 49.01\% & 2.467\\
			MMF3(CoarseG) & 47.96\% & 28.94\% & 58.62\% & 2.577 & 34.46\% & 21.54\% & 48.63\% & 2.426\\
			MMF3 &\textbf{48.92\%} & \textbf{29.99\%} & \textbf{59.97\%} & \textbf{2.683} & \textbf{35.92\%} & \textbf{22.91\%} & \textbf{49.98\%} & \textbf{2.523}\\
			\bottomrule
		\end{tabular}
	\end{table*}
	
	\subsection{Ablation Study}
	
	In order to verify the effectiveness of our proposed FGFM, we conduct ablation experiments for FGFM on JHD and PBD datasets, and the specific experimental results are shown in Table \ref{Tab: table4}.
		
	\textbf{MMF3 (AST) variant:} The embedding vector after fine-grained fusion of tokens embedding and AST embedding in FGFM of MMF3 is replaced with AST embedding. The model is mainly used to verify the effectiveness of tokens embedding in FGFM for code summarization.
		
	\textbf{MMF3 (SAN) variant:} The SAFGF in FGFM of MMF3 is replaced by self-attention, where Q is AST embedding and K and V are tokens embedding. This part mainly verifies that our proposed fine-grained fusion method outperforms the self-attention fusion method.
		
	\textbf{MMF3 (CoarseG) variant:} The SAFGF in the FGFM of MMF3 is replaced with the concatenation of tokens embedding and AST embedding, i.e., a coarse-grained fusion of these two parts. The model is mainly used to verify that our proposed fine-grained fusion method outperforms the coarse-grained fusion method.
	
	Table \ref{Tab: table4} shows the results of the three experiments, and we can observe that MMF3 for multi-modal fine-grained feature fusion achieves the highest scores on the four evaluation metrics. MMF3 (AST) variant has the lowest score among the four methods, mainly because tokens embedding is not used in FGFM, which does not fuse the semantic information of the source code with the structural information of the code represented by the AST. MMF3 (CoarseG) variant scores lower than MMF3 because the coarse-grained fusion of tokens embedding and AST embedding in code snippets in FGFM makes it difficult to learn the rich semantic and syntactic information associated with fine-grained code elements (tokens in sequences and nodes in ASTs) and the associations between them effectively. The MMF3 (SAN) variant scores higher than the MMF3 (CoarseG) variant because multi-head attention can focus on different aspects of tokens embedding and AST embedding and combine the information from all aspects, which is more fully fused than coarse-grained fusion. The MMF3 (SAN) variant scores lower than MMF3 because the multi-head attention only fuses a pair of high-level features compared to the fine-grained fusion method. Thus, the elemental knowledge (token semantics, node attributes) contained in the corresponding modalities and the detailed associations between these fine-grained elements do not receive the explicit attention in fusion, which easily impairs the information related to the code representation. 
	%
	In summary, the crucial components of our proposed fine-grained fusion approach are complementary and indispensable to each other.

	\begin{table}
		\caption{Qualitative example of the different models’ performance on the two datasets}
		\label{Tab: table5}
		\resizebox{1.0\columnwidth}{!}{
			\begin{tabular}{c|l}
				\toprule
				Dataset & Example   \\
				\midrule
				&public static boolean isNumeric (String s)\{\\
				&\quad try\{Double.parseDouble(s);\\
				&\quad return \_BOOL;\\
				&\quad \}catch(Exception e)\{\\
				&\quad return \_BOOL;\}\\
				JHD &\}\\
				&NeuralCodeSum: checks if a string is a valid string.\\
				&MMF3(AST): check if a string have number.\\
				&MMF3(CoarseG):check if a string have a number.\\
				&MMF3((SAN):check if the string is a number.\\
				&MMF3: check if the given string is a number.\\
				&Reference: check if the given string represents a number.\\
				\midrule
				&def get\_image\_file\_path(instance):\\
				&\quad return os.path.join(CONF.instances\_path, instance['name'], 'disk')\\
				&NeuralCodeSum: Return the path of the given instance.\\
				&MMF3(AST): Return the path of the given disk.\\
				PBD &MMF3(CoarseG): Return the path for an instances.\\
				&MMF3((SAN): Generate the path for an instances.\\
				&MMF3: Generate the path for an instances disk.\\
				&Reference: Generate the full path for an instances disk .\\
				\bottomrule
			\end{tabular}
		}
	\end{table}
	
	\subsection{Qualitative Analysis}
	Since the statistics of the experimental results do not visually demonstrate the ability of our proposed method to generate summaries, we take a sample code from each of the two datasets, JHD and PBD. Table \ref{Tab: table5} shows the summaries generated by the different models for the examples in the two datasets. We choose NeuralCodeSum to compare with our model because this baseline is the first official use of the Transformer architecture for the code summarization task. Most of the subsequent models used to do the code summarization task are based on Transformer, so this baseline is representative. Finally, Table \ref{Tab: table5} shows that MMF3 can generate more accurate summaries for code snippets. On the JHD dataset, MMF3 more closely matches the target summary in terms of length and semantic similarity of the generated summary than NeuralCodeSum and these ablation models. For example, the statement generated by MMF3 is different from the reference target summary by only one token and replaces "represents" in the target summary with "is". On the PBD dataset, MMF3 generate more accurate code summary compared to NeuralCodeSum and those ablation models. For example, MMF3 is missing only one token "full" compared to the reference target code summary, while all the other models compared to it have at least two or more tokens different from the reference target summary. The reasons why MMF3 can generate high-quality code summaries are as follows: First, the GCN is used to extract the structural feature information of AST comprehensively. Second, the FGF module is used to realize the fine-grained fusion of the Token modal information of the source code and the AST modal information so that the fused information has both the semantic information of the source code token and the structural information of AST, which provides more clues for generating summaries. Finally, an improved code decoder module is used, which allows the token information of the code and the fused information to be fed into the module to make full use of the input information to predict summaries. As a result, MMF3 can generate higher quality summaries than NeuralCodeSum and other baselines.
	
	\section{RELATED WORK}
	Source code summarization is to generate short natural language descriptions for code snippets. Automatic code summarization approaches can be divided into two main categories, i.e., heuristic/ template-driven approaches and AI/data-driven approaches \cite{leclair2019neural}.
	
	Early approaches to code summarization are mainly heuristic/ template-driven approaches. Haiduc et al. \cite{haiduc2010use,haiduc2010supporting} first coined the term "source code summarization" and proposed a heuristic-based approach, which applied text retrieval techniques to select some important keywords as the generated code comments. After that, many researchers \cite{eddy2013evaluating,sridhara2010towards,sridhara2011automatically,moreno2013automatic,wong2015clocom,mcburney2015automatic,rodeghero2015eye,lu2017learning} designed a set of heuristic rules or created some hand-crafted templates to generate code comments. For example, Sridhara et al. \cite{sridhara2010towards} used the Software Word Usage Model (SWUM) to identify keywords from selected sentences and created summaries.
	
	With the development of deep learning techniques, more and more AI/data-driven approaches are currently proposed for code summarization techniques. Inspired by NMT, some studies treated the code summarization task as a translation task, equivalent to translating code snippets into natural language descriptions. For example, Iyer et al. \cite{iyer2016summarizing} proposed the Seq2Seq model with an attention mechanism. First, code snippets are considered token sequences and input to the encoder. Then the output vector of the encoder is input to the decoder composed of LSTM units. Finally, the corresponding code summaries is output by the decoder. Since the code contains not only the semantic information of the token sequence but also the structural information of the AST, for this reason, more and more researchers have started to apply the structural information of AST to code summarization. Hu et al. \cite{hu2020deep}, Zhang et al. \cite{alon2018code2seq}, and LeClair et al. \cite{leclair2020improved} used various ways to extract information from the AST and input it into the encoder-decoder framework to improve the performance of code summarization. All these studies above used an encoder-decoder framework consisting of LSTM or GRU. In addition, some studies used deep reinforcement learning to enhance code summarization \cite{shido2019automatic,mou2016convolutional,mcburney2014automatic} as the way to address exposure bias during coding \cite{wang2020reinforcement}. Later, with the emergence of the Transformer model, the accuracy of code summarization has been further improved. Ahmad et al. \cite{ahmad2020transformer} proposed the vanilla Transformer and tested it on Java and Python datasets, achieving better results than previous methods. Subsequent approaches to code summarization that have emerged revolve around the Transformer model \cite{wang2021cocosum,zugner2021language,feng2020codebert}.
	
	Compared with the current methods, our proposed MMF3 can fuse the semantic and syntactic structure information of code snippets in a fine-grained way through FGFM and achieve better results than the current state-of-the-art methods.
	
	\section{THREATS TO VALIDITY}
	We have identified the following threats to validity:
	
	\textbf{Fair comparison threat.} Due to hardware limitations \cite{leclair2020improved}, we were unable to conduct fully extensive hyper-parameters optimization for all baselines. This is a problem that deep learning cannot avoid. The hyper-parameters we set when reproducing the work in baselines should match their descriptions as much as possible to mitigate this problem \cite{wan2018improving}.
	
	\textbf{Dataset type.} We used a Java dataset and a Python dataset to validate the effectiveness of our proposed multi-modal fine-grained feature fusion approach to code summarization. However, the code summarization model behaves differently on code snippets written in different languages. Therefore, it is necessary to experiment on larger scope of datasets (e.g., C, C\#) to verify the method's reliability.
	
	\textbf{Evaluation method.} In this paper, we used four evaluation metrics that are more popular in code summarization to assess the similarity between our generated code summarization and the referenced code summarization. Although these evaluation metrics can help us quickly and efficiently calculate the corresponding scores, they cannot provide a comprehensive assessment of the quality of the model-generated code summarization. Therefore, in the future, human evaluation can be considered to be added to the evaluation of the model \cite{yang2021multi}.
	
	\section{CONCLUSION}
	In this paper, we propose a multi-modal fine-grained feature fusion method for neural code summarization. The method characterizes code snippets as two modalities, token sequences and ASTs. It uses FGFM to achieve fine-grained feature fusion of these two modalities, which enables the model to consider the semantic and syntactic structure information of code snippets more comprehensively in generating code summaries than the previous method models, thus improving the accuracy of the generated code summaries.  In addition, we conducted extensive experiments on a Java dataset and a Python dataset. The experimental results demonstrate that our model outperforms the current state-of-the-art model and can effectively improve the accuracy of code summarization. In the future, we plan to add human evaluation to the evaluation metrics of the model. The implementation of MMF3 and the experimental data are publicly available at: \textbf{\url{https://github.com/TransM2/MMF3}}.
	
	\begin{acks}
		This work is financially supported by the Special Project on Innovative Methods (2020IM020100), Natural Science Foundation of Shandong Province, China (ZR2021MF059, ZR2019MF071) and National Natural Science Foundation of China (61602286, 61976127).
	\end{acks}

	\balance	
	\bibliographystyle{ACM-Reference-Format}
	\bibliography{sample-base}
	

\end{document}